\documentclass[lettersize,journal]{IEEEtran}

\usepackage{amsmath}
\usepackage{amssymb}
\usepackage{graphicx} 
\usepackage{subcaption}
\usepackage[utf8]{inputenc}
\usepackage[nolist]{acronym}
\usepackage{booktabs}
\usepackage{booktabs}    
\usepackage[table]{xcolor} 
\usepackage{colortbl}
\usepackage{makecell}
\usepackage{url}

\title{Deep Learning for Virtual Reality User Identification: A Benchmark}

\author{
Davide Frizzo\textsuperscript{1},
Fabrizio Genilotti\textsuperscript{1},
David Petrovic\textsuperscript{1},
Arianna Stropeni\textsuperscript{1},
Francesco Borsatti\textsuperscript{1},\\
Davide Dalle Pezze\textsuperscript{1},
Riccardo De Monte\textsuperscript{1},
Manuel Barusco\textsuperscript{1},
Gian Antonio Susto\textsuperscript{1} \\[6pt]
\textsuperscript{1}University of Padova, Italy \\
\texttt{davide.frizzo.1@phd.unipd.it},
\texttt{fabrizio.genilotti@studenti.unipd.it}, 
\texttt{david.petrovic@phd.unipd.it}, 
\texttt{arianna.stropeni@studenti.unipd.it}, 
\texttt{francesco.borsatti@phd.unipd.it},
\texttt{davide.dallepezze@unipd.it},
\texttt{riccardo.demonte@phd.unipd.it}, 
\texttt{manuel.barusco@phd.unipd.it},
\texttt{gianantonio.susto@unipd.it}
}

\begin{document}
\maketitle

\begin{acronym}
  \acro{VR}{Virtual Reality}
  \acro{HMD}{Head-Mounted Display}
  \acro{IMU}{Inertial Measurement Unit}
  \acro{MLP}{Multi Layer Perceptron}
  \acro{CNN}{Convolutional Neural Network}
  \acro{RNN}{Recurrent Neural Network}
  \acro{LSTM}{Long Short-Term Memory}
  \acro{GRU}{Gated Recurrent Unit}
  \acro{TCN}{Temporal Convolutional Network}
  \acro{SSM}{State Space Models}
  \acro{S4}{Structured State Spaces}
  \acro{S4D}{Diagonal State Space}
  \acro{S5}{Simplified Structured State Space Layers}
  \acro{FFT}{Fast Fourier Transform}
  \acro{SISO}{Single Input Single Output}
  \acro{MIMO}{Multi Input Multi Output}
  \acro{XR}{Extended Reality}
  \acro{BR}{Body Relative}
  \acro{BRV}{Body Relative Velocity}
  \acro{BRA}{Body Relative Acceleration}
  \acro{FLOP}{Floating Point Operations}
  \acro{MRR}{Mean Reciprocal Rank}
  \acro{NLP}{Natural Language Processing}
\end{acronym}

\begin{abstract}

\ac{VR} applications require robust user identification systems to ensure secure access to equipment and protect worker identities. 
Motion tracking data from \ac{VR} headsets and controllers has emerged as a powerful behavioral biometric, with recent studies demonstrating identification accuracies exceeding 94\% across a large user base. However, the application of modern deep learning architectures, particularly \ac{SSM}, to \ac{VR} scenarios remains largely unexplored. 
In this work, we benchmark user identification performance across the large-scale \textit{Who is Alyx} \ac{VR} dataset, gathering data from 71 users playing the popular Half-Life:Alyx game. 
We evaluate both established architectures (\ac{LSTM}, \ac{GRU}, \ac{CNN}, \ac{TCN}, Transformer) and the emerging \ac{SSM}s on time series motion data. 
Our results provide the first comprehensive benchmark of state-of-the-art and novel architectures for \ac{VR} user identification, establishing baseline performance metrics for future privacy-preserving authentication systems in manufacturing environments.
\end{abstract}

\section{Introduction}\label{sec:intro}

The rapid adoption of VR technology across diverse sectors has created new challenges for user authentication and identity protection. Modern VR systems continuously collect rich behavioral data from Head-Mounted Display (HMD) and hand controllers, generating high-dimensional time series that capture individual movement patterns. Recent research has demonstrated that this motion data serves as a powerful biometric identifier, with classification accuracies approaching those of traditional biometric modalities like fingerprints~\cite{nair2023unique}.
\\
Secure user identification in VR environments is critical across multiple application domains. In healthcare, VR-based surgical training and telemedicine platforms require robust authentication to ensure that only qualified medical professionals access patient data and perform remote procedures. In education, virtual classrooms and examination systems need identity verification to maintain academic integrity and prevent credential fraud. 
In manufacturing and industrial settings, access to dangerous equipment, such as forklifts, industrial robots, and machinery, must be restricted to authorized and properly trained personnel, while VR-based training systems serve as certification platforms where identity verification ensures training completion authenticity~\cite{garrido2023sok}.

Furthermore, previous work has predominantly employed traditional machine learning with hand-crafted statistical features or established deep learning architectures (\acp{LSTM}, \acp{CNN}) from the mid-2010s. 
The recent emergence of \acp{SSM} such as \ac{S4}, \ac{S4D} and \ac{S5}~\cite{s4,s4d,s5}, which demonstrate superior performance on long-sequence modeling tasks, has not been systematically evaluated for \ac{VR} user identification.

This paper makes the following contributions:

\begin{itemize}

\item \textbf{Comprehensive benchmarking:} We evaluate user identification
  performance across the \textit{Who is Alyx} dataset using both established deep
  learning architectures and modern \acp{SSM}.

\item \textbf{Extension of feature-based methods:} End-to-end deep learning
  architectures with minimal pre-processing on the raw motion sequences are
  considered, in contrast with previous methods, mainly employing statistical
  feature engineering

\item \textbf{State space model evaluation:} We provide the first systematic
  evaluation of structured state space models (\ac{S4D} and \ac{S5}) for \ac{VR} user
  identification, comparing their performance and computational efficiency
  against traditional architectures.

\item \textbf{Performance-efficiency analysis:} We provide an analysis of accuracy-efficiency trade-offs across architectures, establishing baseline metrics that enable informed model selection for VR authentication systems based on deployment constraints.
\end{itemize}

The remainder of this paper is organized as follows. Section
\ref{sec:related_work} reviews related work in \ac{VR} behavioral biometrics and time series classification.
Section \ref{sec:methodology} describes our methodology, including the VR dataset characteristics (Section \ref{subsec:alyx_dataset}) and the deep learning architectures evaluated in our benchmark (Section \ref{subsec:model_architectures}), with emphasis on SSMs in Section \ref{subsec:ssm}).
Section \ref{sec:exp_setup} describes our experimental setup.
Section \ref{sec:exp_results} presents results and discussion. 
Finally, Section \ref{sec:conclusion} concludes with future research directions.

\begin{figure}[htbp]
    \centering
    \includegraphics[width=0.48\textwidth]{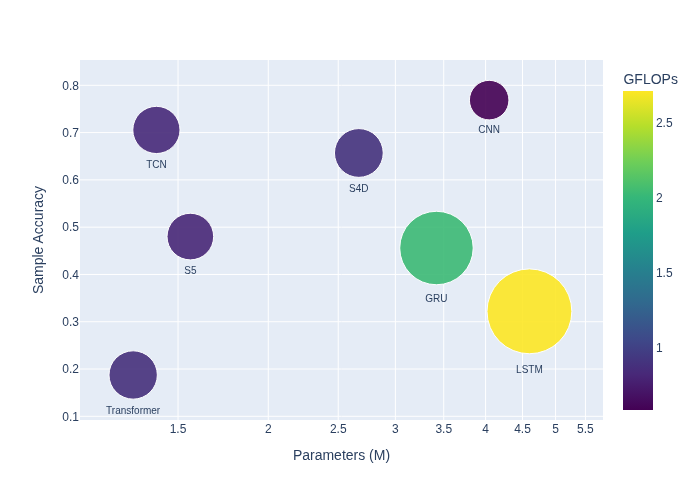}
    \caption{Performance and efficiency trade-offs: comparison of deep learning models across parameter count (millions), sample accuracy, and computational complexity (GFLOPS).}
    \label{fig:performance_efficiency_comparison}
\end{figure}

\section{Related work}\label{sec:related_work}

\subsection{\ac{VR} Behavioral Biometrics for User Identification.}

Behavioral biometrics in VR has progressed from early proofs of concept to scalable identification systems. Pfeuffer et al.~\cite{pfeuffer2019behavioural} demonstrated that body motion patterns and spatial relationships between tracked points (e.g., head–hand distance and inter-hand angles) are discriminative for user identification, showing that movements during basic VR tasks yield user-specific signatures.
Identification research has scaled to large populations with Nair et al.~\cite{nair2023unique}, who analyzed motion data from 55,541 Beat Saber players and achieved up to 94.33\% accuracy using hierarchical classification, with gradient boosting outperforming deep models, indicating that feature-engineered approaches can surpass end-to-end deep learning in this setting.
\\
Despite these advances, behavioral biometrics face temporal stability challenges: Liebers et al.~\cite{liebers2021understanding} showed that body normalization improves cross-session identification by 38\%, but their 2023 field study~\cite{liebers2023exploring} found performance drops from 86\% to 71\% over time, motivating dataset refresh cycles of roughly nine months.
Cross-system generalization presents additional challenges. Miller et al.~\cite{miller2021using} achieved Equal Error Rates between 1.38-7.78\% across different \ac{VR} platforms (Oculus Quest, HTC Vive, Vive Cosmos) using Siamese neural networks. Their subsequent work~\cite{miller2022combining} introduced scale and translation-invariant features that outperformed baselines in 36 of 42 experimental conditions. 
Most recently, Rack et al.~\cite{schach2025motion} demonstrated 78.5\% cross-application identification accuracy using transformer-based similarity learning across five
different \ac{VR} games.
\\
Despite advances in VR behavioral biometrics, most existing studies rely on datasets where each user contributes only a few minutes of recording time.
This data scarcity fundamentally limits the evaluation of deep learning architectures, which require substantial training samples to leverage their capacity for automatic feature learning from raw sequences. 
To properly benchmark deep learning architectures and assess their true potential for VR user identification, we selected the \textit{Who is Alyx}, a publicly available VR biometric dataset, containing a large per-user recording time (approximately 45 minutes per session across two sessions).

\subsection{Time Series Classification Architectures.}

Deep learning for time series classification has progressed through several architectural paradigms. 
The DeepConvLSTM architecture~\cite{ordonez2016deep} established the CNN+LSTM hybrid approach, achieving state-of-the-art results on human activity recognition benchmarks by capturing both spatial features and temporal dynamics. 
InceptionTime~\cite{ismail2020inceptiontime} demonstrated that ensemble Inception networks achieve state-of-the-art performance on the UCR archive while being orders of magnitude faster than previous methods, training on 8 million time series in 13 hours.

Transformer architectures have been adapted for time series, with attention mechanisms enabling long-range dependency modeling. Recent work combines transformers with recurrent units, leveraging attention for global context and GRUs for local temporal patterns. However, transformers face quadratic complexity with respect to the sequence length. 
This poses a critical challenge in extended \ac{VR} sessions.

\ac{SSM} represent an emerging paradigm. The \ac{S4} architecture~\cite{gu2021efficiently} solved the Path-X task (16,000 length sequences) where all prior models failed, with 60× faster generation than transformers. 
These results are attributable to the peculiar architecture of these models which enables parallel training similarly to \ac{CNN} and unbounded context in inference as in \ac{RNN}.

\section{Methodology}
\label{sec:methodology}

\subsection{VR Scenario: Alyx Dataset}
\label{subsec:alyx_dataset}
We evaluate user identification performance on the publicly available \textit{Who is Alyx} \ac{VR} dataset. This dataset~\cite{rack2023alyx} includes tracking data from 71 \ac{VR} users playing Half-Life: Alyx on an HTC Vive Pro across two separate sessions. Each participant played for approximately 45 minutes per session, with sessions conducted on different days and on different chapters of the game to evaluate cross-session identification stability.
\\
\textit{Half-Life: Alyx} provides a rich contextualized environment where users perform diverse actions: navigation via teleportation or physical walking, object manipulation with gravity gloves, combat with weapon reloading, and puzzle solving with hand scanners and 3D puzzles. 
This diversity of interactions makes the dataset particularly challenging for identification, as input sequences contain arbitrary combinations of actions rather than repeated specific movements.
\\
This dataset represents a significant advancement over prior VR behavioral biometric datasets in terms of scale. Each user contributed approximately 90 minutes of gameplay data across two sessions, providing substantially more training data per individual than previous collections. This extended per-user data volume is critical for deep learning approaches, as earlier datasets with shorter recording windows often lacked sufficient samples to effectively train complex neural network architectures.
\\
\textbf{Tracking Modalities:}
The dataset comprises three categories of tracking data: (1) motion data, including position and orientation of the head-mounted display and controllers; (2) gaze data from eye tracking sensors; and (3) physiological signals captured via an Empatica E4 wristband (acceleration, heart rate, electrodermal activity, photoplethysmography, inter-beat interval, and peripheral body temperature) and a Polar H10 chest strap (acceleration and electrocardiogram).
\\
\textbf{Key Challenges:}
This multimodal data presents several key challenges for user identification:
\begin{itemize}
    \item High dimensionality: Modern VR systems track multiple points (e.g., HMD + 2 controllers = 21 features per frame), generating high-dimensional time series.
    \item Long sequences: VR sessions can span tens of minutes at 15-90 Hz sampling rates, producing sequences with tens of thousands of time steps.
    \item Cross-session generalization: Models must identify users from behaviors exhibited in different time periods than the training data, which is challenging due to the fact that user movements can vary across sessions due to task learning effects, fatigue, and environmental factors.
\end{itemize}

\textbf{Temporal Encoding:}
To capture different aspects of user motion patterns, three temporal data
encodings are evaluated, each emphasizing distinct kinematic properties and
providing several levels of feature engineering:

\begin{itemize}
  \item \textbf{\ac{BR} (Body-Relative):} Raw body-relative positions and orientations that preserve the absolute spatial configuration of controllers relative to the HMD reference frame.
  \item \textbf{\ac{BRV} (Body-Relative Velocity):} First-order temporal derivatives representing instantaneous motion dynamics. For positional data, frame-to-frame differences capture the velocity, while quaternion delta rotations encode angular velocity for orientations.
  \item \textbf{\ac{BRA} (Body-Relative Acceleration):} Second-order temporal derivatives computed from \ac{BRV} data, representing acceleration patterns. This encoding captures motion smoothness, jerk characteristics, and rapid directional changes that may reflect individual motor control strategies and reflexes during gameplay.
\end{itemize}

\textbf{Cross-Domain Generalizability:} 
Beyond gaming applications, the Who is Alyx dataset serves as a valuable proxy for evaluating user identification systems across diverse VR domains such as healthcare and manufacturing. 
The variety of motor behaviors captured during gameplay, precision hand movements for object manipulation, coordinated hand-eye actions for targeting, and sustained attention during navigation have the potential to closely mirror the interaction patterns required in professional VR applications. 

Consequently, identification models trained and benchmarked on this dataset provide meaningful insights into their potential performance in safety-critical environments where robust authentication is essential, such as restricting access to virtual surgical platforms or certifying authorized personnel for operating dangerous industrial equipment through VR-based training systems.

\subsection{State Space Models} 
\label{subsec:ssm}

State Space Models represent the latest paradigm shift in sequence modeling
architectures, emerging as a promising alternative to both recurrent and
attention-based approaches. While established architectures like LSTMs, CNNs,
and Transformers have dominated time series classification for the past decade,
SSMs have only recently gained attention in deep learning as an efficient
alternative to attention-based architectures for sequence modeling (\cite{s4}).
Unlike Transformers, whose self-attention mechanism scales quadratically with
sequence length, \ac{SSM}s offer linear complexity, making them particularly
well-suited for modeling long sequences such as the ones recorded in \ac{VR}
sessions. \\ Given their recent emergence and lack of evaluation in the VR
biometrics domain, we dedicate specific attention to SSMs in this benchmark.
Understanding whether these state-of-the-art sequence models can outperform
established architectures on VR motion data have important implications for both
identification accuracy and deployment feasibility on resource-constrained VR
hardware. \\ An \ac{SSM} models the evolution of an internal state $x(t)$
driven by an input signal $u(t)$ and producing an output $y(t)$:

\begin{equation}
  \begin{cases}
      \dot{x}(t) = A x(t) + B u(t) \\
      y(t) = C x(t) + D u(t)
  \end{cases}
\end{equation}

where the system matrices $A$, $B$, $C$, and $D$ are learned from data. 
In their discrete-time formulation, \ac{SSM} closely resembles linear \ac{RNN}, but lacks non-linear activation functions. \\ 
Such a linear structure equips \ac{SSM} with a dual interpretation. On one hand, they can be viewed as
recurrent models that update a hidden state over time, similar to \ac{RNN}. 
On the other hand, unfolding the state dynamics reveals an equivalent convolutional form, where the output is obtained by convolving the input sequence with a fixed kernel. This convolutional view enables parallel training via \ac{FFT}, combining the long-range dependency modeling of \ac{RNN} with the computational efficiency of \ac{CNN}. \\ Several variants of the \ac{SSM} architecture have been proposed over the years~\cite{ssm_survey} across different domains such as \ac{NLP} speech recognition, vision, and time series forecasting. 

However, the most commonly used \ac{SSM} for time series modeling are \texttt{S4}, \texttt{S4D}, and \texttt{S5} (\cite{s4},\cite{s4d},\cite{s5}). \\ The \texttt{S4} model leverages the dual recurrent-convolutional nature of \ac{SSM} and is designed as a \ac{SISO} system, where each input feature is processed independently by a separate \ac{SSM} block and combined through a mixing layer. \texttt{S4D} simplifies the architecture by restricting the system matrices to be diagonal, reducing computational cost at the expense of theoretical expressiveness. 
The \ac{S5} model further simplifies \ac{S4} by adopting a \ac{MIMO} formulation, allowing all input features to be processed jointly within a single \ac{SSM} block.

\subsection{Model Architectures}
\label{subsec:model_architectures}

We conduct a comprehensive benchmark of deep learning architectures spanning several classes of sequence modeling approaches. 
Our evaluation includes traditional recurrent networks (LSTM, GRU), convolutional approaches (CNN, TCN), attention-based models (Transformer), and the recently introduced State Space Models (S4D, S5). 
This diverse architectural comparison enables us to identify which inductive biases, whether convolutional locality, recurrent temporal dynamics, global attention, or structured state space representations, are most effective for capturing discriminative motion patterns in VR behavioral biometrics. 
Furthermore, we are going to analyze the accuracy-efficiency trade-offs achieved by each model, discussing the feasibility of deployment for each model on resource-constrained VR hardware.

\noindent The evaluated architectures are:
\begin{itemize}
    \item Multi-Layer Perceptron (MLP): A feedforward neural network trained on hand-crafted statistical features (min, max, mean, quantiles, and standard deviation). It is used as a baseline to assess the value of end-to-end deep learning approaches versus traditional feature engineering methods.
    \item Convolutional Neural Network (CNN): It applies convolutional filters to extract local temporal patterns from motion sequences, enabling the model to learn motion primitives at different temporal scales.
    \item Long Short-Term Memory (LSTM): A recurrent architecture with gating mechanisms and designed to capture long-term dependencies.
    \item Gated Recurrent Unit (GRU): A simplified recurrent architecture that reduces the computational complexity while maintaining the ability to model temporal dependencies in motion sequences.
    \item Temporal Convolutional Network (TCN): Employs dilated causal convolutions to achieve large receptive fields while maintaining parameter efficiency. The dilated convolutions enable the model to capture long-range temporal patterns without the sequential computation constraints of recurrent models.
    \item Transformer: It utilizes multi-head self-attention mechanisms to model dependencies across the entire input sequence. 
    \item Diagonal State Space (S4D): A structured state space model that restricts system matrices to diagonal form, reducing S4's computational cost while maintaining efficient long-sequence modeling.
    \item Simplified State Space Layer (S5): Further simplifies the SSM architecture by adopting a Multi-Input Multi-Output (MIMO) formulation, allowing all input features to be processed jointly within a single SSM block.
\end{itemize}

\section{Experimental Setup}\label{sec:exp_setup}

\subsection{Alyx Dataset}
The original benchmark study~\cite{rack2023alyx} evaluated on \ac{CNN} and \ac{GRU}
architectures.
Our work extends this benchmark by evaluating other state-of-the-art
approaches, including \ac{SSM}, on the same data.
Since the dataset has two separate sessions for each user, models are trained on session 1 and evaluated on session 2 for all users.

\subsubsection{Data Preprocessing}

The following preprocessing steps were considered:

\begin{itemize}

  \item \textbf{Coordinate System Transformation:} Following Rack et
    al.~\cite{rack2023alyx}, we transform raw scene-relative coordinates to
    body-relative coordinates using the \ac{HMD} as the reference frame. This
    removes positional and absolute orientation information that would allow
    models to overfit to spatial locations rather than movement patterns.

  \item \textbf{Downsampling to 15 fps}: Since the majority of the sessions
    are recorded at 15 fps we decided to keep this as the sampling frequency
    of the time series signals by downsampling the data from the 90 fps
    sessions.

  \item \textbf{Removal of noisy samples}: The first and last minute of each
    session are removed to discard noisy samples.

\end{itemize}

Each encoding results in 18 features per frame: 7 features per controller (position x,y,z; rotation quaternion x,y,z,w) plus 4 features for HMD rotation (pitch and yaw only; roll and position become constant in body-relative coordinates).

\subsubsection{Sequence Windowing}

We segment continuous recordings into fixed-length input sequences. 
In particular, we use 20-second windows (300 frames at 15 Hz), consistent with the original benchmark.
Sequences are sampled with 50\% overlap during training to increase the dataset size. 
During evaluation, we use non-overlapping windows and apply majority voting across all windows in a test recording to produce a single user prediction per session.

\subsection{Training Configuration}

\begin{figure*}[h!] 
    \centering
    \begin{subfigure}[b]{0.31\textwidth}
        \centering
        \includegraphics[width=\textwidth]{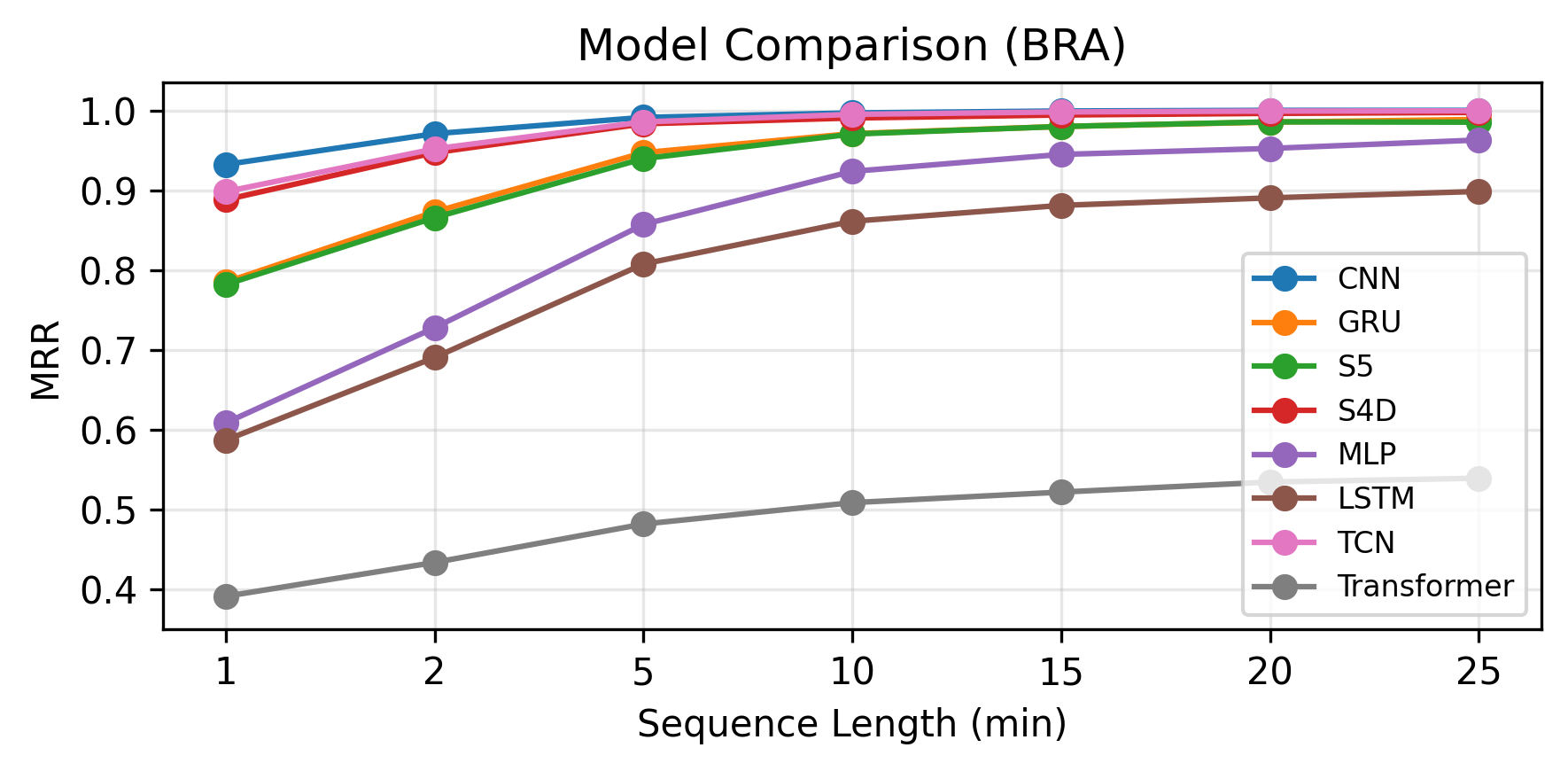}
        \caption{\ac{BRA}}
        \label{fig:bra-acc-plot}
    \end{subfigure}
    \hfill 
    \begin{subfigure}[b]{0.31\textwidth}
        \centering
        \includegraphics[width=\textwidth]{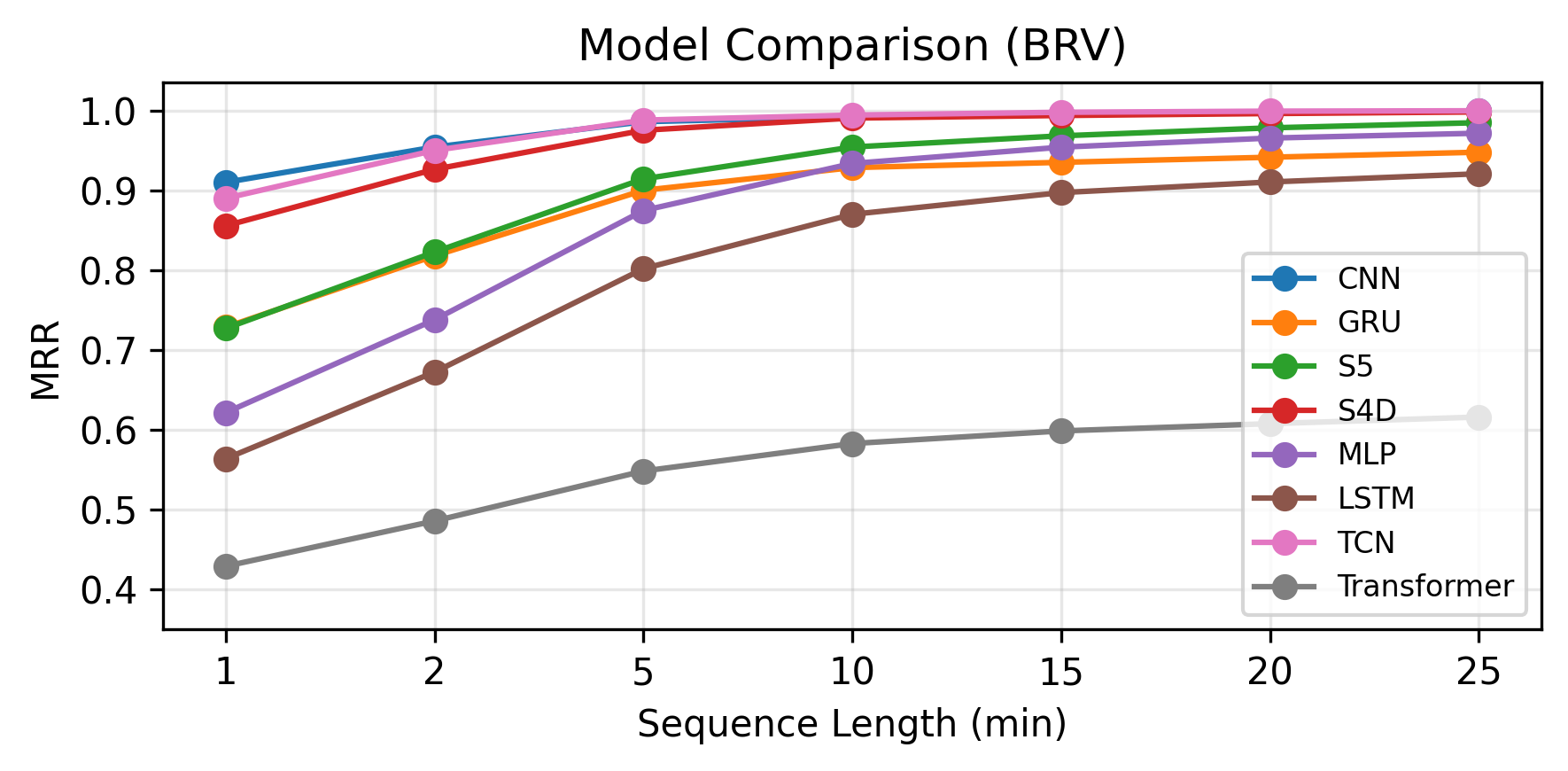}
        \caption{\ac{BRV}}
        \label{fig:brv-acc-plot}
    \end{subfigure}
    \hfill 
    \begin{subfigure}[b]{0.31\textwidth}
        \centering
        \includegraphics[width=\textwidth]{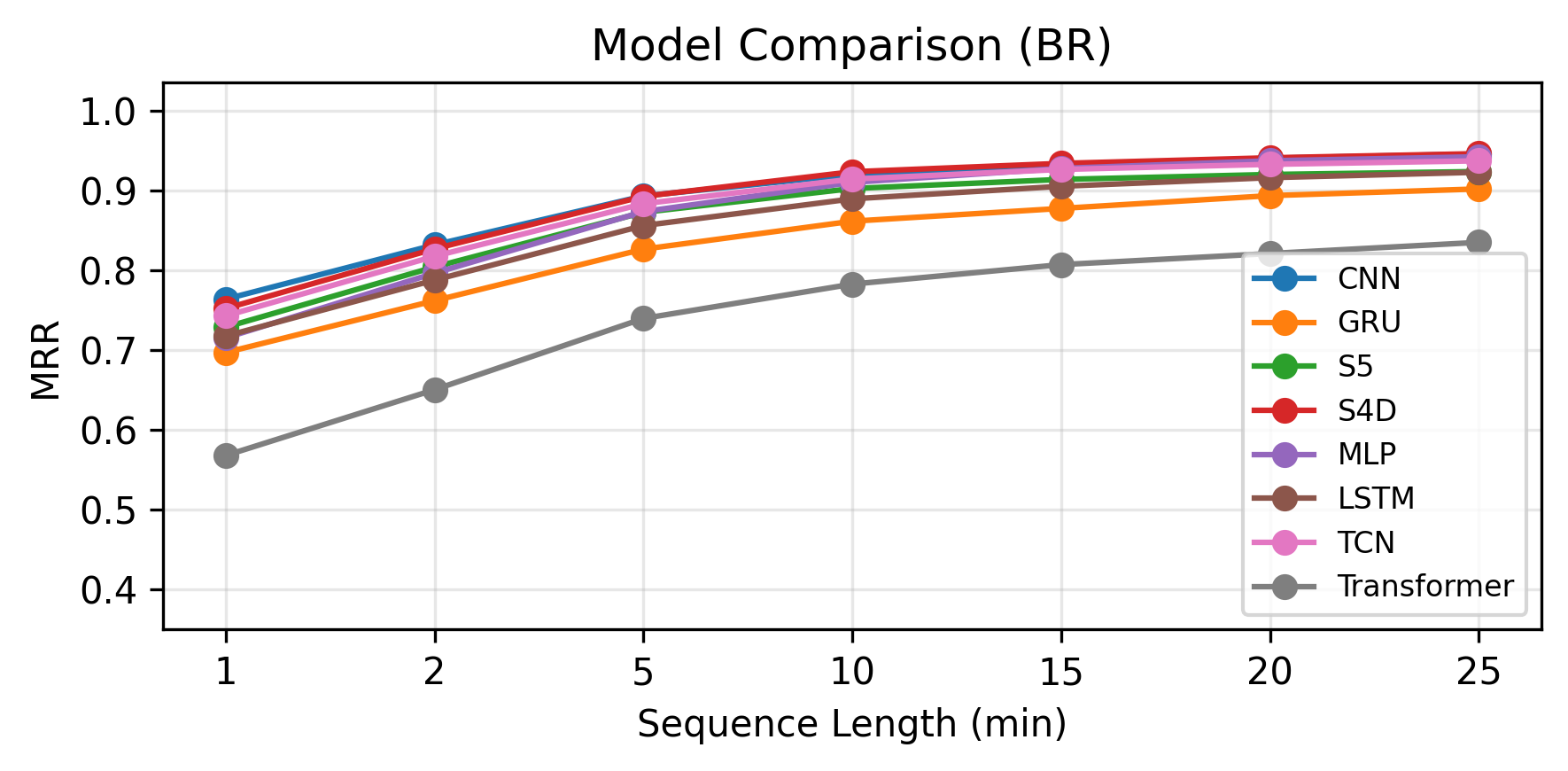}
        \caption{\ac{BR}}
        \label{fig:br-acc-plot}
    \end{subfigure}

    \caption{Test sample size vs. \ac{MRR} for the \ac{BRA}, \ac{BRV} and \ac{BR} encodings}
    \label{fig:acc-plot}
\end{figure*}

Regarding the training configuration, the Adam optimizer is used with
Early Stopping and a patience factor of 5 to mitigate overfitting. In
particular, the set of model's parameters used for the inference evaluation is
chosen by minimizing the mean accuracy on the validation set (i.e., the last 5
minutes of a training session).

\subsection{Evaluation Metrics}

\noindent \textbf{Computation Complexity: }
\ac{FLOP}s are considered to measure the computational complexity of a model as they provide a rough estimate of a model's training and inference time.
\\
\textbf{Performance: } Taking into account the difficulty of the task, the \ac{MRR} metric, is considered at this stage. 
MRR quantifies the ranking quality of the model’s predictions by measuring, for each sequence, the position at which the correct class appears in the ordered list of predicted classes. 
For each sequence window, the classes are ranked according to their predicted scores, and the reciprocal of the rank of the ground-truth class is computed. 
The final MRR is obtained by averaging these reciprocal ranks over all sequences, providing a continuous measure of how consistently the correct class appears near the top of the ranked predictions. 
\\
\textbf{Memory: }
Another considered metric in the study is the number of trainable parameters, since it is directly related to the memory required during the inference phase.
Memory is a critical parameter when considering the deployment of DL models on standalone VR headsets with limited onboard memory.

\begin{table}[h!]
    \centering
    \begin{tabular}{l c c}
        \toprule
        Model         & Parameters (M) & FLOPs (GFLOPs)  \\
        \midrule
        \texttt{MLP}  & 0.8935     & 0.0018     \\
        \texttt{CNN}  & 4.046     & 0.5886     \\
        \texttt{GRU}  & 3.42     & 2.0324     \\
        \texttt{LSTM}  & 4.6     & 2.7101     \\
        \texttt{TCN}  & 1.4     & 0.8394     \\
        \texttt{Transformer} & 1.3 & 0.8793     \\
        \texttt{S4D} & 2.67     & 0.8951   \\
        \texttt{S5}  & 1.56 & 0.818        \\
        \bottomrule
    \end{tabular}
    \caption{Model parameters and Mult-Adds operations}
    \label{tab:model_params}
\end{table}

\begin{figure}[htbp]
    \centering
    \includegraphics[width=0.4\textwidth]{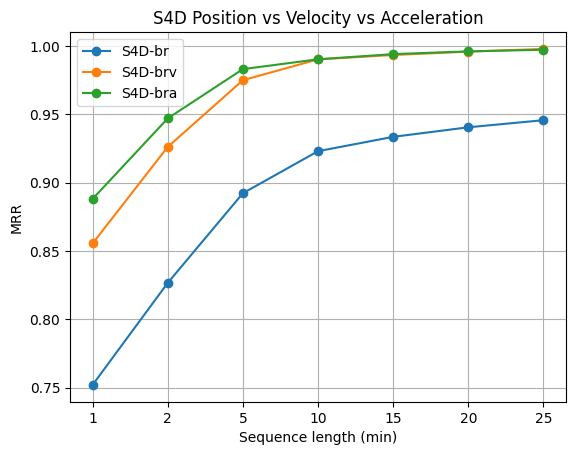}
    \caption{Test sample size vs. \ac{MRR}}
    \label{fig:s4d-plot}
\end{figure}

\section{Results}\label{sec:exp_results}

In this section, we highlight the experimental results obtained on the Who is
Alyx dataset.

\subsection{Performance Comparison Across Architectures}

First, we evaluate the capabilities of deep learning models in \ac{VR} User
Identification.

In Figure \ref{fig:acc-plot} the \ac{MRR} metric is evaluated in relation to
the length of the time-series presented to the model at test time, showing the
results based on the used encoding among \ac{BRA}, \ac{BRV}, and \ac{BR}
represented in \ref{fig:bra-acc-plot}, \ref{fig:brv-acc-plot}, and
\ref{fig:br-acc-plot} respectively.

The first observation is that the \ac{MRR} increases as the amount of test data
increases for all data encodings, as majority voting across longer sequences
reduces prediction noise and improves classification stability on the sequence.
Moreover, comparing the curves associated with the different models, it is
possible to conclude that \ac{CNN},\ac{TCN}, and \ac{S4D} are the most accurate
ones, while \ac{LSTM} and Transformer are at the bottom of the ranking. In
particular, the Transformer model performs significantly worse than the other
architectures. 
This result is likely due to the limited dataset size and absence of large-scale pretraining.
This gives the intuition that the continuous time nature of the \ac{VR} sensor readings is more appropriate for models like \ac{TCN} and \ac{SSM}.

Finally, the \ac{MLP} model shows inferior results, placed between the
best-performing models and the \ac{LSTM} model. This result confirms the
superiority of end-to-end deep learning approaches using sequential models
rather than extracting statistical features from raw data. This analysis is
confirmed in Figure \ref{fig:s4d-plot}, where the performances of the \ac{S4D}
model on the three different data encodings are compared.

\subsection{Impact of Temporal Encoding Strategies}

Among the three data encodings considered, the \ac{BRA} encoding produced the
best performances across all models in accordance with~\cite{rack2023alyx},
followed by \ac{BRV} and \ac{BR} (as visible in Figure \ref{fig:s4d-plot} for the S4D model). In particular, the \ac{BRA} and \ac{BRV} show the most significant differences in \ac{MRR} in the low test data regimes and
saturate to overlapped results for sequences longer than 10 minutes. 
On the
other hand, the \ac{BR} encoding shows significantly worse results. 
In fact, velocity and acceleration information, obtained through consecutive sample
differences on the raw position and velocity data, emphasize motion dynamics
over absolute spatial positioning. These derivative-based features are
invariant to constant offsets and linear drift in the tracking system,
capturing how users move, which proves more discriminative for behavioral identification.

\subsection{Memory and Computational Efficiency Considerations for Deployment} %

To conduct a final evaluation not only in terms of model performances but also
with respect to efficiency and computational requirements, Figure
\ref{fig:performance_efficiency_comparison} captures all three metrics in a
single plot and shows the trade-offs across architectures. In particular, the
horizontal axis contained the amount of model parameters (in millions), the
model's sample accuracies in the \ac{BRA} data encoding are reported in the
vertical axis, and the size of the blob's radius is proportional to the model's
\ac{FLOP}s. \\
Examining the performance axis, convolutional-based architectures (CNN, TCN)
and SSMs like S4D emerge as the top-performing models, significantly
outperforming recurrent (LSTM, GRU) and attention-based (Transformer)
approaches.
In addition, the top three models (CNN, TCN, S4D) share similar computational
requirements (FLOPs), indicating comparable training and inference times.
However, CNN's high parameter count (4.0M) makes it memory-bound, potentially
limiting deployment on standalone VR headsets with constrained RAM. TCN (1.4M
parameters) and S4D (2.7M parameters) offer a better choice under this aspect:
they achieve only marginally lower accuracy than CNN while requiring 65\% and
33\% fewer parameters, respectively. \\ \textit{Therefore, TCN and S4D represent optimal choices when deployment constraints are considered, achieving
competitive performance with substantially lower memory footprints.}

\section{Conclusion}\label{sec:conclusion}

This study provided a comprehensive benchmark of deep learning architectures for user identification within \ac{VR} contexts. By evaluating a spectrum of models ranging from traditional \acp{RNN} to modern \acp{SSM} on
the Who is Alyx dataset, several key performance metrics and trade-offs were
established.

The experimental results demonstrate that \ac{BRA} encoding consistently
provides the most discriminative features for identification across all
architectures. Performance analysis reveals that \ac{CNN}, \ac{TCN}, and the
\ac{S4D} model represent the state-of-the-art for this task, significantly
surpassing Transformers and recurrent models. Notably, while \acp{CNN} achieve
high accuracy, their memory requirements may be prohibitive for certain
applications. In contrast, \ac{TCN} and \ac{S4D} architectures offer a superior
balance between identification accuracy, parameter efficiency, and
computational demand, making them highly suitable for deployment on standalone
\ac{VR} hardware.

The robust user identification capabilities demonstrated in this benchmark have significant practical implications across multiple domains. In healthcare settings, reliable VR-based authentication can ensure that only qualified medical professionals access surgical training platforms and telemedicine systems. In industrial environments, accurate user identification enables secure access control to dangerous equipment such as industrial robots and forklifts. The performance-efficiency trade-offs established in this work can inform the deployment of authentication systems on resource-constrained VR hardware in real-world scenarios.

Future research directions include addressing cross-application generalization challenges for \acp{SSM}, studying the impact of user learning and adaptation on long-term identification accuracy, and developing efficient model compression techniques for edge deployment.

\bibliographystyle{IEEEtran}  
\bibliography{main}

\end{document}